\documentclass[lettersize,journal]{IEEEtran}

\usepackage{cite}
\usepackage{amsmath,amssymb,amsfonts}
\usepackage{algorithmic}
\usepackage{graphicx}
\usepackage{textcomp}
\usepackage{xcolor}
\usepackage{CJKutf8}
\usepackage{tabularray}
\usepackage{subfigure}
\usepackage{caption}
\usepackage{multirow}
\usepackage{bm}
\usepackage{cite}
\usepackage{amsmath,amssymb,amsfonts}
\usepackage{algorithmic}
\usepackage{graphicx,color}
\usepackage{textcomp}
\usepackage{xcolor}
\usepackage{hyperref}
\usepackage{booktabs}
\usepackage{algorithm,algorithmic}
\usepackage{graphicx} 
\def\BibTeX{{\rm B\kern-.05em{\sc i\kern-.025em b}\kern-.08em
    T\kern-.1667em\lower.7ex\hbox{E}\kern-.125emX}}

\begin{document}



\title{
Entropy-and-Channel-Aware Adaptive-Rate Semantic Communication with MLLM-Aided Feature Compensation
}

\author{Weixuan Chen,~\IEEEmembership{Graduate Student Member, IEEE}, Qianqian Yang,~\IEEEmembership{Member, IEEE}, Yuhao Chen,~\IEEEmembership{Student Member, IEEE}, Chongwen Huang, Qian Wang,~\IEEEmembership{Member, IEEE}, Zehui Xiong, Zhaoyang Zhang,~\IEEEmembership{Senior Member, IEEE}



\thanks{ 

This paper was presented partially in {IEEE} {GLOBECOM}, Kuala Lumpur, Malaysia, Dec. 2023 \cite{chen2023deep}.

Weixuan Chen, Qianqian Yang$^{\dag}$, Yuhao Chen, Chongwen Huang, and Zhaoyang Zhang are with the College of Information Science and Electronic Engineering, Zhejiang University, Hangzhou 310027, China. 
(e-mails: \{weixuanchen, qianqianyang20$^{\dag}$, csechenyh, chongwenhuang, ning\_ming\}@zju.edu.cn). 

Qian Wang is with the Institute of Cyberspace Security, Zhejiang University of Technology, Hangzhou 310023, China. (e-mail: wangqian18@zjut.edu.cn).

Zehui Xiong is with the School of Electronics, Electrical Engineering and Computer Science, Queen's University Belfast, Belfast, BT7 1NN, U.K. (e-mail: z.xiong@qub.ac.uk).

This work is partly supported by the NSFC under grant No. 62293481, No. 62571487, No. 62201505, by the National Key R\&D Program of China under Grant 2024YFE0200802, and by the Zhejiang Provincial Natural Science Foundation of China under Grant No. LZ25F010001.  (Corresponding author: Qianqian Yang.)

}
}

\maketitle

\begin{abstract}

Despite the transmission efficiency gains of semantic communication (SemCom) over traditional methods, most existing SemCom schemes still operate at a fixed transmission rate regardless of channel conditions and transmitted content, resulting in wasted resources in favorable channels and degraded performance in harsh channels.
To address this issue, we propose a novel SemCom framework that incorporates an entropy-and-channel-aware adaptive rate control mechanism over MIMO Rayleigh fading channels.
Specifically, we embed a joint representation of the channel state information (CSI) and the signal-to-noise ratio (SNR) into both the semantic encoder and decoder, thereby realizing channel-aware semantic coding and decoding.
Moreover, the proposed method jointly exploits the CSI, the SNR, the feature maps, and their 2D entropy via two policy networks to selectively transmit only a subset of feature maps and, within each selected feature map, only a subset of symbols. 
Thereby, it achieves finer-grained adaptive rate control than existing methods. 
%
At the receiver, leveraging the strong visual understanding capability of multimodal large language models (MLLMs), we deploy the lightweight visual encoder (InternViT-300M) of the pre-trained InternVL3.5 model to compensate for discarded feature maps and symbols, and we fine-tune InternViT using low-rank adaptation (LoRA) for parameter-efficient training.
Experimental results show that, with a carefully designed channel-aware loss function, our system automatically allocates more communication resources under poor channels to enhance task performance while reducing resource usage under favorable channels and maintaining high task performance. 
Our approach consistently outperforms both conventional separation-based source and channel coding and state-of-the-art (SOTA) adaptive-rate SemCom methods in terms of rate-distortion performance, achieving about 0.4-0.9 dB higher PSNR than the SOTA adaptive-rate method at similar compression ratios.
\end{abstract}




\section{Introduction}

In recent years, semantic communication (SemCom) has gained significant attention as a promising alternative communication paradigm with the potential to surpass the traditional Shannon capacity limit \cite{luo2022semantic,liu2022indirect}.
SemCom \cite{10766058,chen2026securedigitalsemcom} enhances bandwidth efficiency by selectively extracting and transmitting only the crucial information relevant to specific transmission tasks, i.e., \textit{semantic information}, while discarding non-essential content. 
This makes SemCom an attractive solution for wireless communication applications that generate large volumes of data traffic.
Existing SemCom approaches typically leverage advanced deep learning techniques to extract semantic information from the source data at the transmitter and to reconstruct the source data at the receiver through end-to-end training.
These approaches have demonstrated excellent performance in transmitting various data types, including text \cite{han2022semantictext,peng2022robust}, speech \cite{han2022semantic1,weng2023deep}, images/videos \cite{han2023generative,chen2024nearly,tang2024contrastive,chen2024enhancing,chi2025deepguard}, and multimodal data \cite{luo2022multi,xie2022task,wan2023cooperative}.


However, most existing SemCom methods \cite{farsad2018deep,xie2021deep,zhang2022wireless,han2023generative} directly map source data to channel input symbols without explicitly modeling the task-dependent importance of different symbols.
This uniform treatment overlooks the varying significance of transmitted symbols for the downstream task, thereby missing the opportunity to improve task performance through more judicious allocation of communication resources.
Moreover, many SemCom systems adopt a fixed neural network and a fixed transmission rate for coding and decoding, which limits their adaptability to varying channel conditions and prevents them from fully exploiting available communication resources to enhance task performance.

%


Regarding importance-aware SemCom systems,
Liu \emph{et al.} \cite{liu2024ofdm} proposed a semantic importance measurement method for OFDM-based SemCom systems, incorporating both feature-task correlations and inter-feature correlations to dynamically allocate more reliable subcarriers to higher-priority semantic features.
Gao \emph{et al.} \cite{gao2024importance} introduced a metric, termed semantic value, to measure the importance of semantic features for text transmission based on Zipf's distribution, where word frequency influences semantic value. 
Liang \emph{et al.} \cite{2025liangsemanticimp} proposed a semantic-importance-aware MIMO SemCom framework that learns unequal importance levels of semantic symbols via bilateral progressive training and exploits them for importance-aware eigenmode mapping and power allocation through singular value decomposition (SVD)-based precoding.
Overall, these studies prioritize important semantic features to improve task performance and communication efficiency.

Several studies have also explored multi-rate or adaptive-rate SemCom systems.
For example, 
Kurka \emph{et al.} \cite{kurka2019successive, kurka2021bandwidth} proposed bandwidth-agile deep joint source-channel coding (DeepJSCC) schemes that encode an image into multiple layered codewords for successive refinement and multiple descriptions, enabling reconstruction from different subsets of received layers, with reconstruction quality improving as more layers become available.
Bian \emph{et al.} \cite{bian2023deepjscc} investigated bandwidth- and signal-to-noise ratio (SNR)-adaptive DeepJSCC, where the channel SNR and bandwidth ratio are fed into the model as side information to optimize performance under different channel SNRs and transmission rates. 
Luo \emph{et al.} \cite{2025luoadmit} proposed ADMIT for one-to-many image transmission, enabling a single model to adapt to different bandwidth ratios and channel SNRs through latent-channel prioritization and bandwidth-aware truncation, together with an SNR-aware decoder, while requiring no channel awareness at the encoder.
Other approaches, such as entropy-based rate control, have been developed to adaptively select symbols according to their semantic content. 
For instance, Bao \emph{et al.} \cite{2025baomdvsc} proposed the MDVSC framework, which employs entropy-based semantic importance coding to discard low-entropy symbols under bandwidth constraints, enabling explicit and precise control of code length while maintaining communication quality.


Furthermore, 
Yang \emph{et al.} \cite{yang2022deep} designed a policy-network-based scheme that automatically adjusts the number of transmitted feature groups according to both the channel SNR and the image content.
Zhang \emph{et al.} \cite{zhang2023predictive} proposed a predictive and adaptive coding framework that predicts reconstruction quality from the channel conditions, compression ratio, and image content, and then selects the optimal compression ratio under a target quality constraint to automatically set the coding rate.
Shi \emph{et al.} \cite{2024shiimage} built a cross-attention-based probabilistic graph to construct a hierarchical semantic parse tree and performed multi-level variable-length coding on semantic feature nodes, assigning longer codewords to patches with stronger semantic connectivity and shorter ones to less important regions, thus enabling content-aware adaptive coding rates under a given bandwidth constraint.
Yang \emph{et al.} \cite{2025swinjscc} proposed SwinJSCC, which employs the Swin Transformer as the codec backbone and introduces channel and rate adaptation modules. The latter takes a pre-specified target rate as input and learns binary masks over latent channels to select informative components, enabling a single model to support adaptive-rate wireless image transmission across different bandwidth ratios.
For multimodal tasks, He \emph{et al.} \cite{he2023rate} proposed a multimodal SemCom framework with rate-adaptive coding, where the semantic importance of each modality is defined by its noise sensitivity and coding rates are assigned accordingly to reduce inference delay.
Additional studies can be found in \cite{wang2022wireless,dai2022nonlinear,zhou2022adaptive,huang2023toward,xie2023memory,dai2023towardadaptive,gao2023adaptive,zhu2023semantic,gong2023adaptive,gao2024rateadaptive,jiang2024large,lyu2024semantic,yang2025rateadaptive}.
The above works mainly consider SISO SemCom systems. 
Meanwhile, several recent studies have explored multi-rate or adaptive-rate strategies \cite{yao2023learnedMIMO,2025scscmimo} for MIMO SemCom systems.

Nevertheless, existing studies on importance-aware and multi-rate/adaptive-rate SemCom still exhibit several limitations.
First, many approaches allocate communication resources based on manually defined importance metrics, which may not fully capture the true task relevance of semantic features. Allowing neural networks to automatically learn how to assess feature importance and allocate resources accordingly has the potential to yield more effective strategies.
Second, some adaptive-rate SemCom methods only select those features that are globally important for the task, but overlook the fact that even within an important feature map, a considerable portion of symbols may be semantically redundant and thus removable.
Third, existing methods rarely consider explicit receiver-side compensation for features discarded at the transmitter, which could otherwise enhance task performance.
Moreover, several multi-rate SemCom systems are trained under fixed channel conditions. When the actual channel state deviates significantly from the training setting, this mismatch can cause non-negligible performance degradation.
Finally, the transmission rates in many existing multi-rate or adaptive-rate systems \cite{kurka2019successive,kurka2021bandwidth,bian2023deepjscc,2025swinjscc,gao2023adaptive,wang2022wireless} are restricted to a few predetermined discrete values rather than any continuous value within a range, resulting in limited rate flexibility.




To overcome the limitations of existing multi-rate or adaptive-rate SemCom approaches, we propose a novel SemCom system with entropy-and-channel-aware adaptive rate control over MIMO Rayleigh fading channels.
Specifically, we incorporate several channel condition adaptive modules (CCAMs) into both the semantic encoder and decoder.
These modules modulate the feature maps based on the current feature maps and a joint embedding of the channel state information (CSI) and the SNR, enabling the system to perform adaptive coding and decoding under varying channel conditions.
In addition, we introduce two policy networks to achieve fine-grained rate adaptation. The first policy network performs feature map selection by discarding unnecessary feature maps, while the second prunes the retained feature maps by removing semantically redundant symbols.
The policy networks take as input the feature maps, their 2D entropy \cite{liu2023filter}, the CSI, and the SNR to make transmission decisions.
Moreover, since multimodal large language models (MLLMs) possess strong visual understanding capabilities \cite{2024mllm}, we employ the visual encoder (InternViT-300M) of the pre-trained InternVL3.5-1B model \cite{internvl35} at the receiver to explicitly compensate for discarded feature maps and symbols, thereby further enhancing task performance. To reduce the training overhead, we fine-tune InternViT using low-rank adaptation (LoRA) \cite{lora2022}.


The main contributions of this paper are as follows:

$\bullet$ \textit{Entropy-and-Channel-Aware Adaptive Rate Control}: 
We propose a novel entropy-and-channel-aware adaptive rate control scheme that enables the semantic encoder and decoder to adapt to varying channel conditions. 
Moreover, the proposed method jointly exploits the feature maps, their 2D entropy, the CSI, and the SNR to determine which feature maps and symbols should be transmitted, effectively reducing redundancy while maintaining high task performance under diverse conditions.


$\bullet$ \textit{Fine-Grained Joint Feature Map Selection and Pruning}: 
Unlike most existing methods that only select task-relevant feature maps, our approach further removes redundancy at the symbol level.
We design two specialized policy networks that are conditioned on the image content, the CSI, and the SNR. They first select informative feature maps and then prune semantically unimportant symbols within them. 
By generating the selection and pruning masks via thermometer encoding, only a single cut-off index for the pruning mask needs to be specified at the receiver, incurring negligible overhead.



$\bullet$ \textit{Channel-Aware Rate-Semantic Tradeoff}:
We design a channel-aware rate-distortion loss that couples semantic task performance with transmission rate (channel usage) across varying channel conditions.
By imposing different penalties on channel usage at different channel conditions, the proposed loss encourages the system to allocate more channel resources under poor channels to enhance task performance, while saving unnecessary resources under favorable channels without sacrificing task performance.



$\bullet$ \textit{MLLM-Aided Feature Compensation}:
We leverage a pre-trained MLLM with strong visual understanding capabilities to explicitly compensate for discarded features and symbols. In particular, we use the MLLM's visual encoder to recover and denoise the received feature maps and symbols, bringing them closer to the original complete representations that are obtained before selection and pruning, and adopt an efficient fine-tuning strategy to adapt the visual encoder to our task. This MLLM-aided feature compensation further enhances overall task performance.

Experimental results show that the proposed system consistently outperforms conventional separation-based source and channel coding schemes as well as state-of-the-art (SOTA) adaptive-rate SemCom methods in terms of rate-distortion performance, achieving about 0.4-0.9 dB higher peak signal-to-noise ratio (PSNR) than the SOTA adaptive-rate method at similar compression ratios.



\section{System Model}

\begin{figure}[t]

\begin{center}
\centerline{\includegraphics[width=1\linewidth]{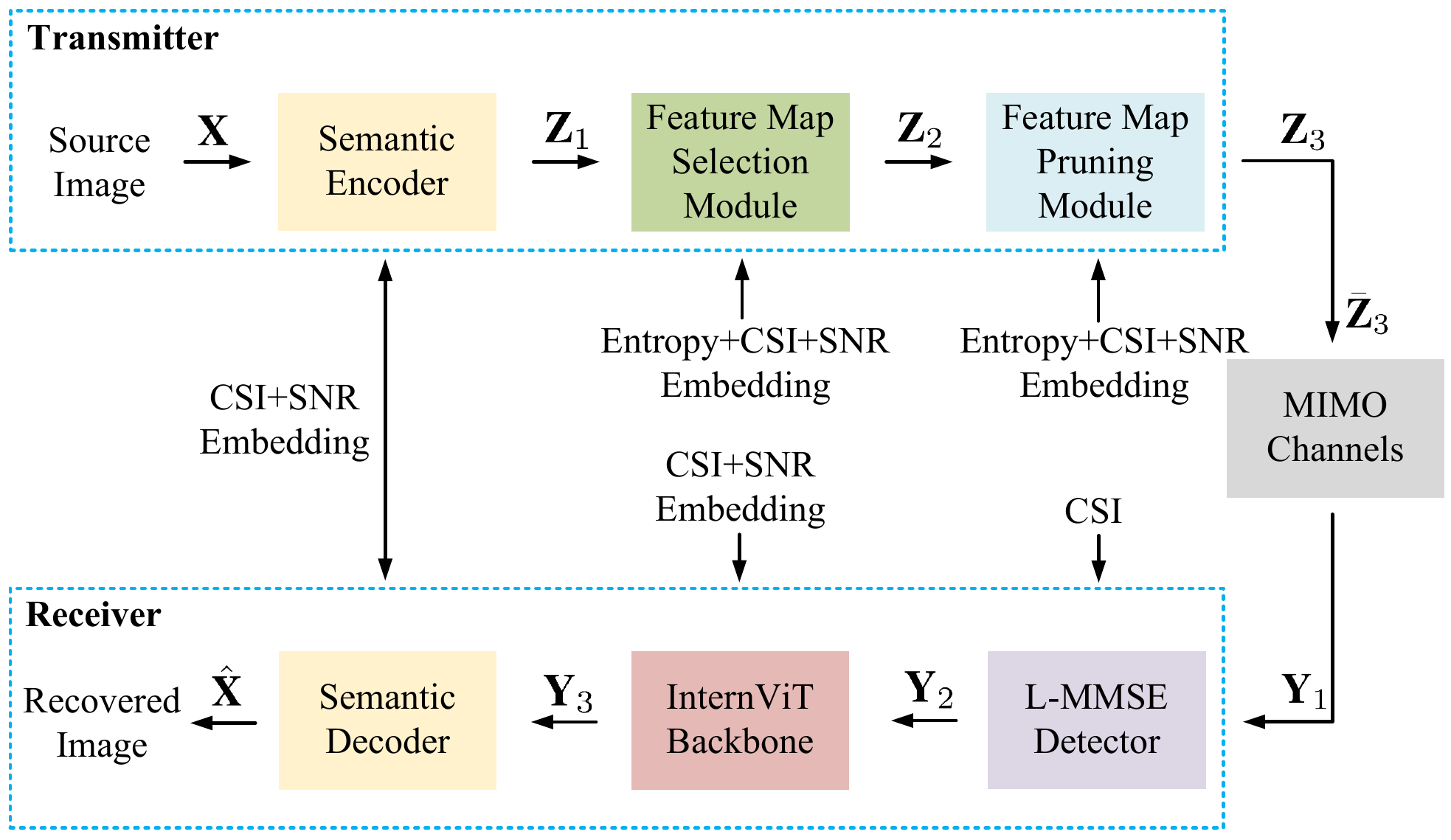}}
\caption{The overall architecture of the proposed SemCom system.}
\label{fig.1}
\end{center}
\vskip -0.3in
\end{figure}

In this paper, we consider the image SemCom problem in an $N_t \times N_r$ MIMO uplink scenario, with a transmitter equipped with $N_t$ antennas and a receiver equipped with $N_r$ antennas. 
The transmitter Alice aims to transmit an image $\textbf{X}$ to the receiver Bob through the MIMO Rayleigh fading channels, as shown in Fig.~\ref{fig.1}.

The transmitter consists of a semantic encoder, a feature map selection module, and a feature map pruning module.  
The semantic encoder exploits both the source image and the channel-condition embedding, which is a joint embedding of the CSI and the SNR, to extract channel-adaptive feature maps, denoted by
\begin{equation}
\textbf{Z}_{1} = f_{\mathrm{se}}(\textbf{X},\hat{\textbf{CH}}_{\rm emb};\bm{\theta}^{\mathrm{se}}),
\end{equation}
where $f_{\mathrm{se}}(\cdot)$ represents the semantic encoder, $\bm{\theta}^{\mathrm{se}}$ refers to its learnable parameters, $\textbf{X}$ is the image to be transmitted, $\hat{\textbf{CH}}_{\rm emb}$ is the channel-condition embedding, and $\textbf{Z}_{1}$ denotes the corresponding output feature maps.
Notably, $\hat{\textbf{CH}}_{\rm emb}$ is obtained by
\begin{equation}
    \hat{\textbf{CH}}_{\rm emb} = f_{\mathrm{ce}}([\operatorname{vec}(\hat{\textbf{H}}), \hat{\mathrm{SNR}}];\bm{\theta}^{\mathrm{ce}}),
\end{equation}
where $f_{\mathrm{ce}}(\cdot)$ denotes the channel-state encoder implemented by a two-layer multilayer perceptron (MLP), $\bm{\theta}^{\mathrm{ce}}$ refers to its learnable parameters, $\hat{\textbf{H}} \in \mathbb{C}^{N_r \times N_t}$ is the estimated MIMO channel matrix (CSI) whose $(i,j)$-th entry represents the channel gain between the $j$-th transmit antenna and the $i$-th receive antenna, and $\hat{\mathrm{SNR}}$ is the estimated average received SNR of the link. 
Considering a conventional setting, we assume $N_t = N_r$.

The pair $(\hat{\textbf{H}},\hat{\mathrm{SNR}})$ constitutes the estimated channel condition, denoted by $\hat{\textbf{CH}}$, used in this work. 
In practice, we first reshape $\hat{\textbf{H}}$ into a real-valued vector $\operatorname{vec}(\hat{\textbf{H}})$, concatenate it with $\hat{\mathrm{SNR}}$, and then feed the resulting $(2N_r N_t + 1)$-dimensional real-valued vector into the MLP to obtain $\hat{\textbf{CH}}_{\rm emb}$.
To acquire $\hat{\textbf{CH}}$, the transmitter first sends a sequence of known pilot symbols. Based on the received pilots, the receiver performs channel estimation to obtain $\hat{\textbf{H}}$ and estimates the average received SNR $\hat{\mathrm{SNR}}$, and then feeds these estimates back to the transmitter. The length of the pilot sequence is set to $1/16$ of the number of semantic information symbols. The detailed procedures for obtaining $\hat{\textbf{H}}$ and $\hat{\mathrm{SNR}}$ will be discussed later.

Then the feature maps $\textbf{Z}_{1}$, together with the entropy-and-channel-condition (EC) embedding, are fed into the feature map selection module, which adaptively selects a subset of important feature maps for transmission, i.e.,
\begin{equation}
    \textbf{Z}_{2} = f_{\mathrm{fms}}(\textbf{Z}_{1}, \textbf{EC}_{\rm emb}; \bm{\theta}^{\mathrm{fms}}),
\end{equation}
where $f_{\mathrm{fms}}(\cdot)$ denotes the feature map selection module parameterized by $\bm{\theta}^{\mathrm{fms}}$.
$\textbf{Z}_{2}$ is the selected subset of feature maps that will be transmitted, where all non-selected feature maps are set to zero.
$\textbf{EC}_{\rm emb}$ is the entropy-and-channel-condition embedding that encodes the CSI, the SNR, and the information content of $\textbf{Z}_{1}$.

The EC embedding $\textbf{EC}_{\rm emb}$ is obtained by first computing the 2D entropy \cite{liu2023filter} of each feature map in $\textbf{Z}_{1}$ and aggregating them into a per-sample entropy descriptor $\textbf{EN}$, and then fusing $\textbf{EN}$ with the estimated channel condition $\hat{\textbf{CH}}$ via a two-layer MLP, i.e.,
\begin{equation}
    \textbf{EC}_{\rm emb} = f_{\mathrm{ec}}([\hat{\textbf{CH}}, \textbf{EN}]; \bm{\theta}^{\mathrm{ec}}),
\end{equation} 
where $f_{\mathrm{ec}}(\cdot)$ denotes the entropy-and-channel-condition encoder with parameters $\bm{\theta}^{\mathrm{ec}}$.

Afterward, the selected feature maps $\textbf{Z}_{2}$ and $\textbf{EC}_{\rm emb}$ are further processed by the feature map pruning module, which adaptively prunes the symbol dimension of each selected feature map. Specifically, this operation can be written as
\begin{equation}
    \textbf{Z}_{3} = f_{\mathrm{fmp}}(\textbf{Z}_{2}, \textbf{EC}_{\rm emb}; \bm{\theta}^{\mathrm{fmp}}),
\end{equation}
where $f_{\mathrm{fmp}}(\cdot)$ denotes the feature map pruning module parameterized by $\bm{\theta}^{\mathrm{fmp}}$, and $\textbf{Z}_{3}$ denotes the resulting pruned feature maps where the chosen symbols remain unchanged and all pruned symbols are set to zero. 

Then, the feature maps $\textbf{Z}_{3}$ are rearranged into a transmit symbol matrix, converted to a complex-valued representation, and normalized to satisfy an average power constraint $P$. 
We denote the normalized feature maps by $\bar{\textbf{Z}}_{3} \in \mathbb{C}^{N_t \times \frac{T}{2}}$. 
The normalized feature maps are then transmitted over the $N_t \times N_r$ MIMO Rayleigh fading channel, and the received signal at the receiver is given by
\begin{equation}
    {\textbf{Y}}_{1} = \textbf{H}\,\bar{\textbf{Z}}_{3} + \textbf{N},
\end{equation}
where ${\textbf{Y}}_{1} \in \mathbb{C}^{N_r \times \frac{T}{2}}$ denotes the received feature maps, 
$\textbf{H} \in \mathbb{C}^{N_r \times N_t}$ is the MIMO channel matrix, and $\textbf{N} \in \mathbb{C}^{N_r \times \frac{T}{2}}$ is the additive white Gaussian noise (AWGN) matrix. 
The channel coefficients follow an i.i.d. circularly symmetric complex Gaussian distribution $h_{i,j} \sim \mathcal{CN}(0,\sigma_h^{2})$ corresponding to Rayleigh fading, and the noise samples are i.i.d. circularly symmetric complex Gaussian $\mathcal{CN}(0,\sigma_n^{2})$, with $\sigma_n^{2}$ determined by the target SNR.

The channel SNR can be calculated as
\begin{equation}
    \mathrm{SNR}
    = \frac{\|\textbf{H}\,\bar{\textbf{Z}}_{3}\|_{F}^{2}}
           {\|\textbf{H}\,\bar{\textbf{Z}}_{3}\|_{0}\,\sigma_{n}^{2}},
    \label{snr-eq}
\end{equation}
where $\|\cdot\|_{F}$ denotes the Frobenius norm. 
This definition corresponds to the average received SNR per nonzero received symbol.
However, since the SNR is required in our encoding process, we need to estimate it beforehand using the transmission of pilot signals. 
To achieve this, we employ the least squares (LS) algorithm to obtain the estimated MIMO channel matrix $\hat{\textbf{H}}$ based on the received pilot signals. 
Specifically, a block of known pilot symbols collected in the matrix $\textbf{P}_{\mathrm{I}}\in\mathbb{C}^{N_t\times T_p}$ is transmitted, and the corresponding received pilot signal at the receiver can be written as
\begin{equation}
    \textbf{Y}_{p} = \textbf{H}\,\textbf{P}_{\mathrm{I}} + \textbf{N}_{p},
\end{equation}
where $\textbf{Y}_{p}\in\mathbb{C}^{N_r\times T_p}$ denotes the received pilot matrix and 
$\textbf{N}_{p}\in\mathbb{C}^{N_r\times T_p}$ is the AWGN matrix. 
Based on $\textbf{Y}_{p}$ and the known pilot matrix $\textbf{P}_{\mathrm{I}}$, the LS estimator yields an estimate $\hat{\textbf{H}}$ of the true MIMO channel matrix. 
The SNR is then estimated as
\begin{equation}
    \hat{\mathrm{SNR}}
    = \frac{\|\hat{\textbf{H}}\,\textbf{P}_{\mathrm{I}}\|_{F}^{2}}
           {\|\hat{\textbf{H}}\,\textbf{P}_{\mathrm{I}}\|_{0}\,\sigma_{n}^{2}},
\end{equation}
where $\sigma_{n}^{2}$ is assumed to be known to the receiver.  
Then, $\hat{\mathrm{SNR}}$, together with the estimated CSI $\hat{\textbf{H}}$, is sent back to the transmitter.

The receiver consists of a linear minimum mean square error (L-MMSE) detector, an InternViT backbone, and a semantic decoder.
The L-MMSE detector is employed to recover an estimate of the transmitted symbols from the received feature maps $\textbf{Y}_{1}$.  
Using the estimated CSI $\hat{\textbf{H}}$ and the noise variance $\sigma_{n}^{2}$, the L-MMSE detection can be written as
\begin{equation}
    \tilde{\textbf{Y}}_{2}
    =
    \left(
        \hat{\textbf{H}}^{H}\hat{\textbf{H}}
        + \sigma_{n}^{2}\textbf{I}_{N_t}
    \right)^{-1}
    \hat{\textbf{H}}^{H}
    \textbf{Y}_{1},
\end{equation}
where $\textbf{I}_{N_t}$ denotes the $N_t\times N_t$ identity matrix, and $\tilde{\textbf{Y}}_{2}\in\mathbb{C}^{N_t\times \frac{T}{2}}$ is the L-MMSE estimate of the transmitted symbol matrix, having the same dimension as $\bar{\textbf{Z}}_{3}$.
Then, $\tilde{\textbf{Y}}_{2}$ is rearranged back into the real-valued feature map format, yielding $\textbf{Y}_{2}$ with the same dimension as $\textbf{Z}_{3}$.

To further mitigate channel distortion and compensate for discarded feature maps and symbols, we introduce an InternViT-based feature compensation module.
Its inputs are the distorted and partially received feature maps $\textbf{Y}_{2}$ and the channel-condition embedding $\hat{\textbf{CH}}_{\rm emb}$, and its output is the refined feature maps $\textbf{Y}_{3}$, formally given by
\begin{equation}
    \textbf{Y}_{3}
    = f_{\mathrm{vit}}(\textbf{Y}_{2}, \hat{\textbf{CH}}_{\rm emb}; \bm{\theta}^{\mathrm{vit}}),
\end{equation}
where $f_{\mathrm{vit}}(\cdot)$ denotes the InternViT-based feature compensation module parameterized by $\bm{\theta}^{\mathrm{vit}}$.
In our design, $f_{\mathrm{vit}}(\cdot)$
is trained to produce $\textbf{Y}_{3}$ that is as close as possible to the original feature maps $\textbf{Z}_{1}$.
This feature refinement helps compensate for the performance loss caused by channel impairments and adaptive feature selection and pruning.

Finally, the refined feature maps $\textbf{Y}_{3}$ are fed into the semantic decoder together with the channel-condition embedding to reconstruct the source image.
The semantic decoder is largely symmetric to the encoder and performs channel-aware semantic decoding, which can be expressed as
\begin{equation}
    \hat{\textbf{X}}
    = f_{\mathrm{sd}}(\textbf{Y}_{3}, \hat{\textbf{CH}}_{\rm emb}; \bm{\theta}^{\mathrm{sd}}),
\end{equation}
where $f_{\mathrm{sd}}(\cdot)$ denotes the semantic decoder parameterized by $\bm{\theta}^{\mathrm{sd}}$, and $\hat{\textbf{X}}$ is the reconstructed image.

We use the PSNR as the performance metric to evaluate the fidelity of the reconstructed images, which is defined as
\begin{equation}
    \mathrm{PSNR} = 10 \log_{10}\left( \frac{\mathrm{MAX}^{2}}{\mathrm{MSE}} \right),
\end{equation}
where $\mathrm{MAX}$ represents the maximum pixel value of the source image (255 in this paper), and $\mathrm{MSE}$ denotes the mean squared error between the source image and the reconstructed image.

The number of \textit{real-valued} channel symbols used to transmit one image is denoted by $S$, and the size of the source image is $H \times W \times 3$. 
The effective compression ratio (CR) of the proposed system is defined as
\begin{equation}
    CR = \frac{S}{2 \times 3 H W}.
\end{equation}
The objective of our proposed system is to optimize image reconstruction performance while minimizing the required compression ratio under varying channel conditions.

\section{Proposed Method}

In this section, we present the proposed SemCom framework in detail. We first describe the channel-aware semantic encoder and decoder, then introduce the joint feature map selection and pruning module for entropy-and-channel-aware adaptive rate control, followed by the InternViT-based feature compensation module. Finally, we discuss the design of our channel-aware multi-objective loss function.

\subsection{Semantic Encoder and Decoder}

\subsubsection{Semantic Encoder}

\begin{figure}[t]
\begin{center}
\centerline{\includegraphics[width=1\linewidth]{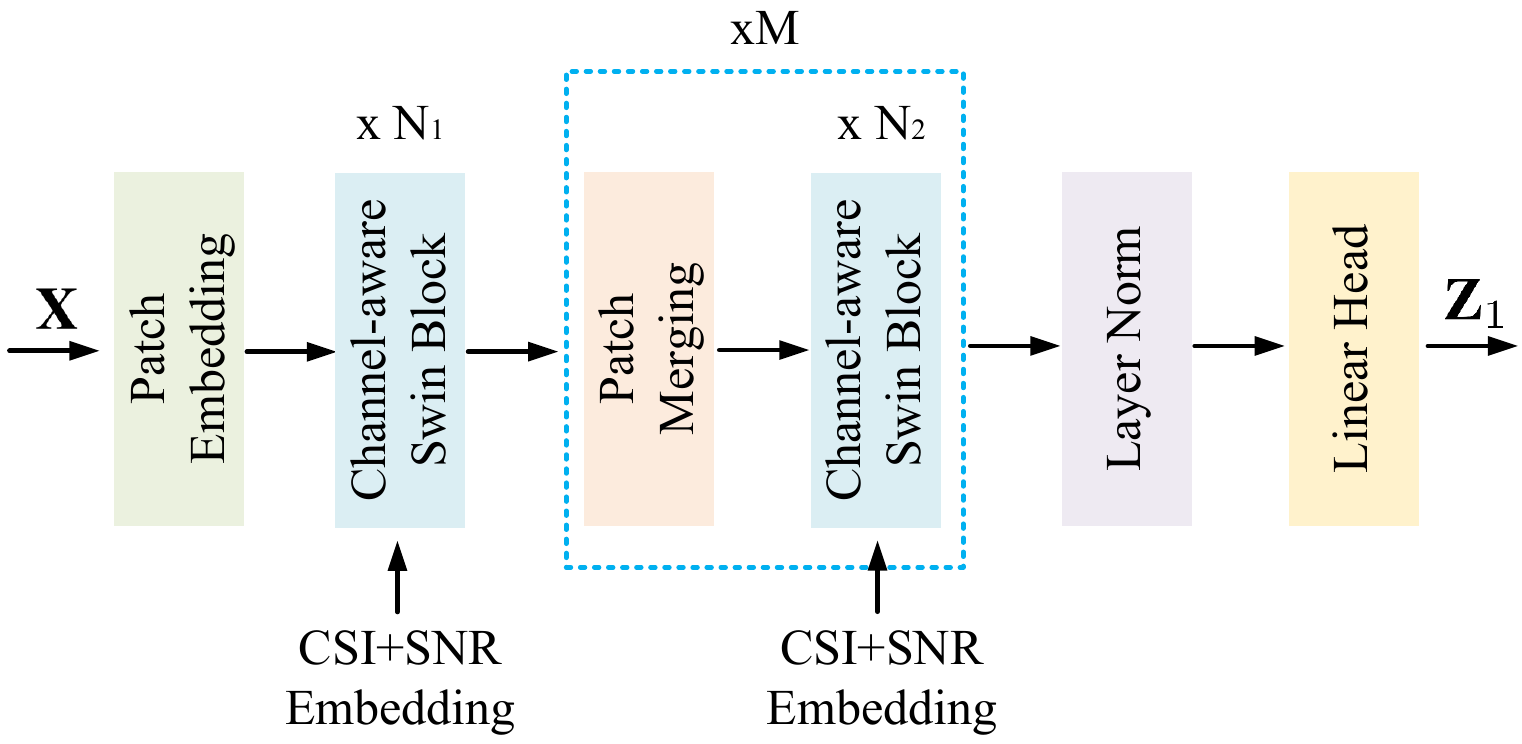}}
\caption{The network architecture of the semantic encoder.}
\label{fig.se}
\end{center}
\vskip -0.3in
\end{figure}

\begin{figure}[t]

\begin{center}
\centerline{\includegraphics[width=1\linewidth]{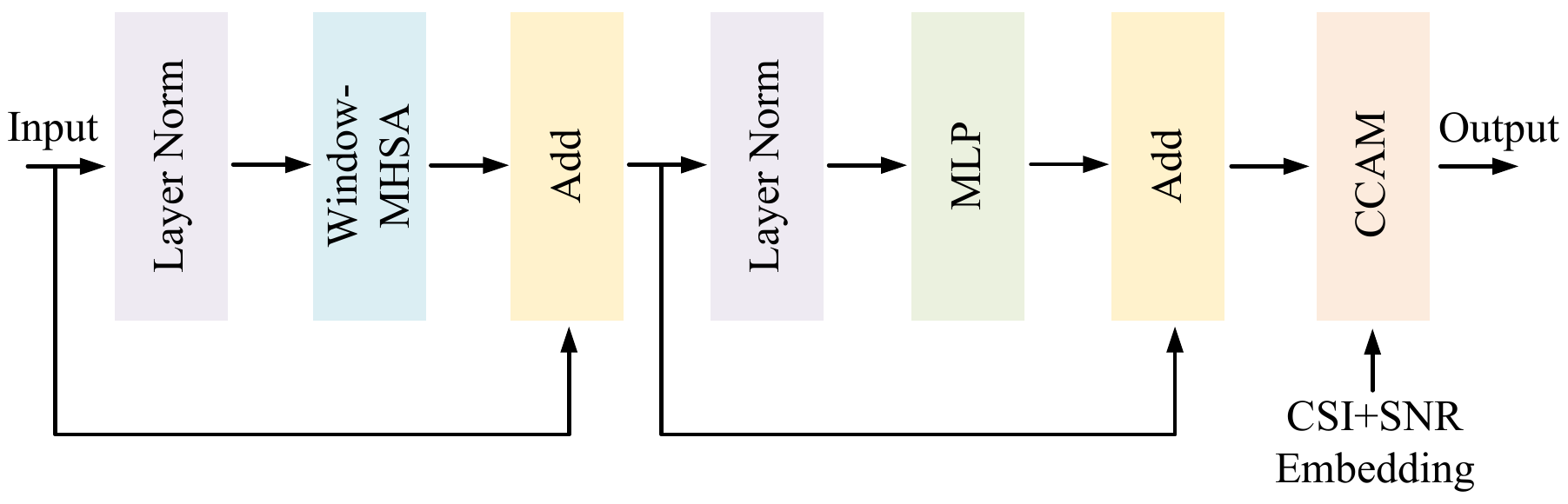}}
\caption{The network architecture of the channel-aware Swin Transformer block.}
\label{fig.swinblock}
\end{center}
\vskip -0.3in
\end{figure}

\begin{figure}[t]

\begin{center}
\centerline{\includegraphics[width=1\linewidth]{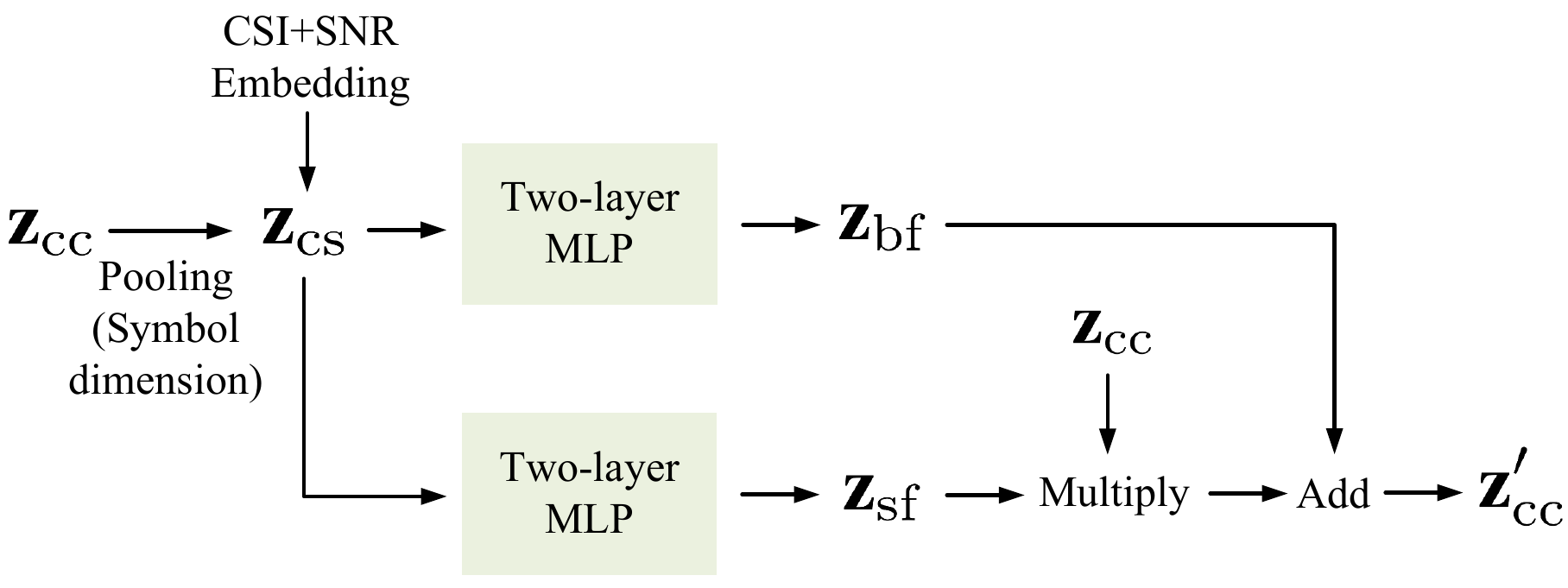}}
\caption{The network architecture of the channel condition adaptive module (CCAM). }

\label{fig.ccam}
\end{center}
\vskip -0.3in
\end{figure}

\begin{figure}[t]

\begin{center}
\centerline{\includegraphics[width=0.85\linewidth]{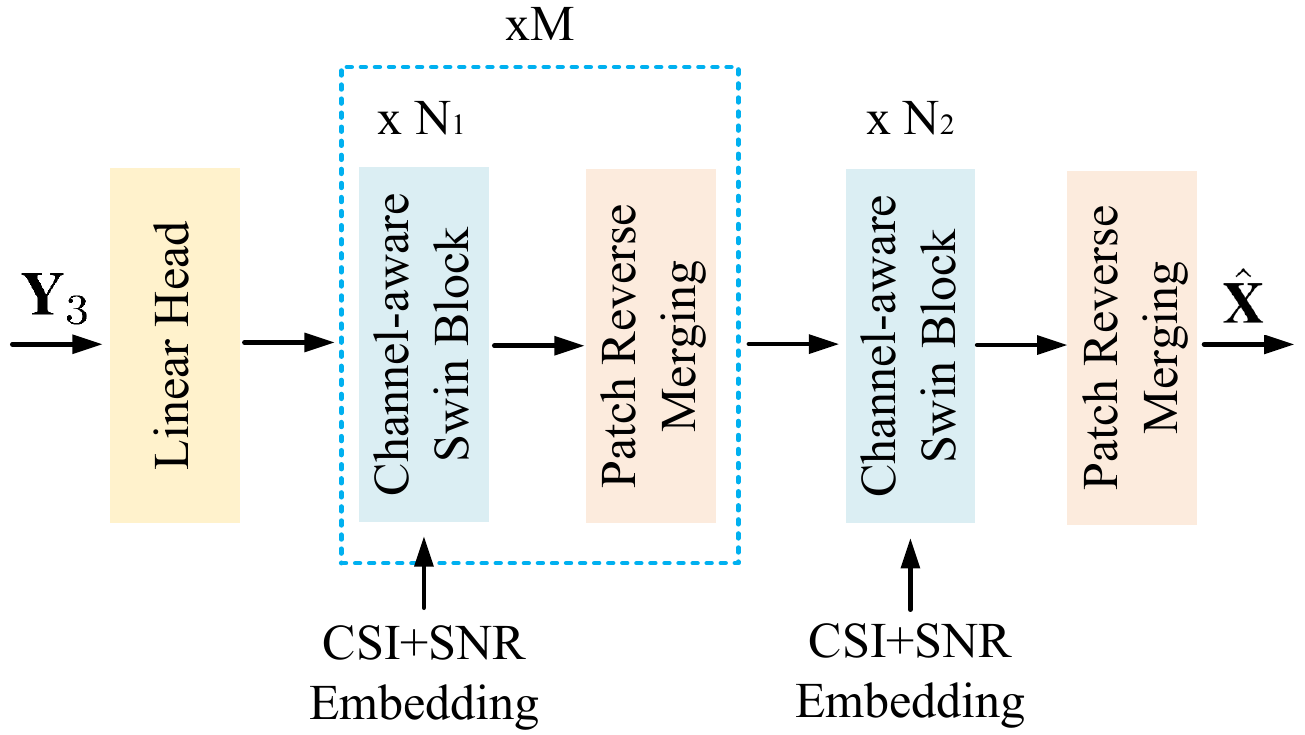}}
\caption{The network architecture of the semantic decoder.}
\label{fig.sd}
\end{center}
\vskip -0.3in
\end{figure}

As illustrated in Fig.~\ref{fig.se}, the semantic encoder maps the input RGB image $\textbf{X}\in\mathbb{R}^{H\times W \times 3}$ into a compact latent representation while taking the channel conditions into account.
The encoder is designed based on the SwinJSCC architecture \cite{2025swinjscc}, but each Swin Transformer block is extended to be channel-aware via the channel-condition embedding $\hat{\textbf{CH}}_{\rm emb}$.
Specifically, the encoder is organized into a stack of channel-aware Swin Transformer stages.
The first stage takes the input image, divides it into non-overlapping patches, applies a patch embedding layer to project each patch into a low-dimensional token in the feature space (a 1D feature map along the symbol dimension), and then processes the resulting token sequence with $N_1$ channel-aware Swin Transformer blocks.
Each block employs window-based multi-head self-attention (W-MHSA).
The subsequent stages consist of patch-merging layers that reduce the spatial resolution while increasing the channel (feature map) dimension, together with several channel-aware Swin Transformer blocks ($N_2, N_3, \ldots$).
As a result of the hierarchical patch-merging operations, the spatial resolution of the feature maps is gradually reduced (e.g., from $H\times W$ to $\frac{H}{2}\times\frac{W}{2}$, then to $\frac{H}{4}\times\frac{W}{4}$, and so on at each downsampling stage), while the channel dimension is increased accordingly.
To enable adaptation to the channel conditions, $\hat{\textbf{CH}}_{\rm emb}$ is provided as a global conditioning vector to each channel-aware Swin Transformer block and is used to modulate the intermediate features within the block.
After passing through all stages, a layer normalization is applied, followed by a linear projection layer that maps the final channel dimension to a fixed size $CU$.
The resulting feature maps are denoted by $\textbf{Z}_{1}$ and have spatial dimension $\frac{H}{2^{i}}\times \frac{W}{2^{i}}$ and channel dimension $CU$, where $i$ is the number of downsampling stages (i.e., patch-merging operations).
Equivalently, after flattening the spatial dimensions, $\textbf{Z}_{1}$ can be viewed as an $L\times CU$ latent representation, where $L = \frac{H}{2^{i}}\times \frac{W}{2^{i}}$ is the symbol dimension, and $CU$ is a user-defined channel dimension that controls the overall compression ratio.

\subsubsection{Channel-Aware Swin Transformer Block}

As illustrated in Fig.~\ref{fig.swinblock}, each channel-aware Swin Transformer block takes as input the feature maps and the channel-condition embedding, and outputs refined feature maps that are adapted to the current channel conditions.
The overall structure follows the standard Swin Transformer block with residual connections, augmented by an additional channel condition adaptive module (CCAM) at the end.
Given the input feature maps, the block first applies layer normalization, followed by a window-based multi-head self-attention module with optional window shifting.
The attention output is then added to the input through a residual connection.
Subsequently, the features are processed by another layer normalization and an MLP, whose output is again added to the input of this sublayer via a second residual connection.
To incorporate channel awareness, the block further includes a CCAM that takes both the intermediate feature maps and the channel-condition embedding as inputs and produces the final output of the block.
The CCAM exploits the channel-condition embedding to modulate the channel-wise activations of the feature maps, thereby adapting the features to current channel conditions.

\subsubsection{Channel Condition Adaptive Module}

The network architecture of the proposed channel condition adaptive module is illustrated in Fig.~\ref{fig.ccam}. 
Let $\textbf{z}_{\mathrm{cc}}\in\mathbb{R}^{L_{\rm z}\times C_{\rm z}}$ denote the input feature maps to this module, where $L_{\rm z}$ is the symbol dimension and $C_{\rm z}$ is the channel dimension.
First, a global average pooling is applied to $\textbf{z}_{\mathrm{cc}}$ along the symbol dimension to obtain a channel-wise summary vector $\bar{\textbf{z}}_{\mathrm{cc}}\in\mathbb{R}^{C_{\rm z}}$.
This vector is then concatenated with the channel-condition embedding, yielding
$\textbf{z}_{\mathrm{cs}} = [\,\bar{\textbf{z}}_{\mathrm{cc}},\,\hat{\textbf{CH}}_{\rm emb}\,]$.
The vector $\textbf{z}_{\mathrm{cs}}$ is fed into two separate two-layer MLPs to produce a channel-wise scaling factor $\textbf{z}_{\mathrm{sf}}$ and a channel-wise bias factor $\textbf{z}_{\mathrm{bf}}$, respectively.
Finally, the input feature maps are modulated via channel-wise affine transformation as
\[
    \textbf{z}_{\mathrm{cc}}'
    = \textbf{z}_{\mathrm{cc}} \odot \textbf{z}_{\mathrm{sf}} + \textbf{z}_{\mathrm{bf}},
\]
where $\odot$ denotes element-wise multiplication. 
$\textbf{z}_{\mathrm{sf}}$ and $\textbf{z}_{\mathrm{bf}}$ are broadcast along the symbol dimension to match the shape of $\textbf{z}_{\mathrm{cc}}$.
$\textbf{z}_{\mathrm{cc}}'$ is the output of the CCAM.
By injecting the channel-condition embedding into the computation of the scaling and bias factors, the CCAM enables the model to adaptively emphasize or suppress different feature channels under varying channel conditions.

\subsubsection{Semantic Decoder}

As illustrated in Fig.~\ref{fig.sd}, the semantic decoder is largely symmetric to the encoder.
It takes the refined feature maps $\textbf{Y}_{3}$ as input and first applies a linear head to map the channel dimension $CU$ back to the decoder feature dimension.
Then, in each decoder stage, several channel-aware Swin Transformer blocks are followed by a patch reverse-merging layer, which gradually increases the spatial resolution while reducing the channel dimension.
The channel-condition embedding $\hat{\textbf{CH}}_{\rm emb}$ is injected into all channel-aware Swin Transformer blocks in the same way as in the encoder.
After the final stage, the feature maps are reshaped and projected to three channels to obtain the reconstructed image $\hat{\textbf{X}}\in\mathbb{R}^{H\times W \times 3}$.

\subsection{Joint Feature Map Selection and Pruning Module}

\begin{figure}[t]
\begin{center}
\centerline{\includegraphics[width=0.95\linewidth]{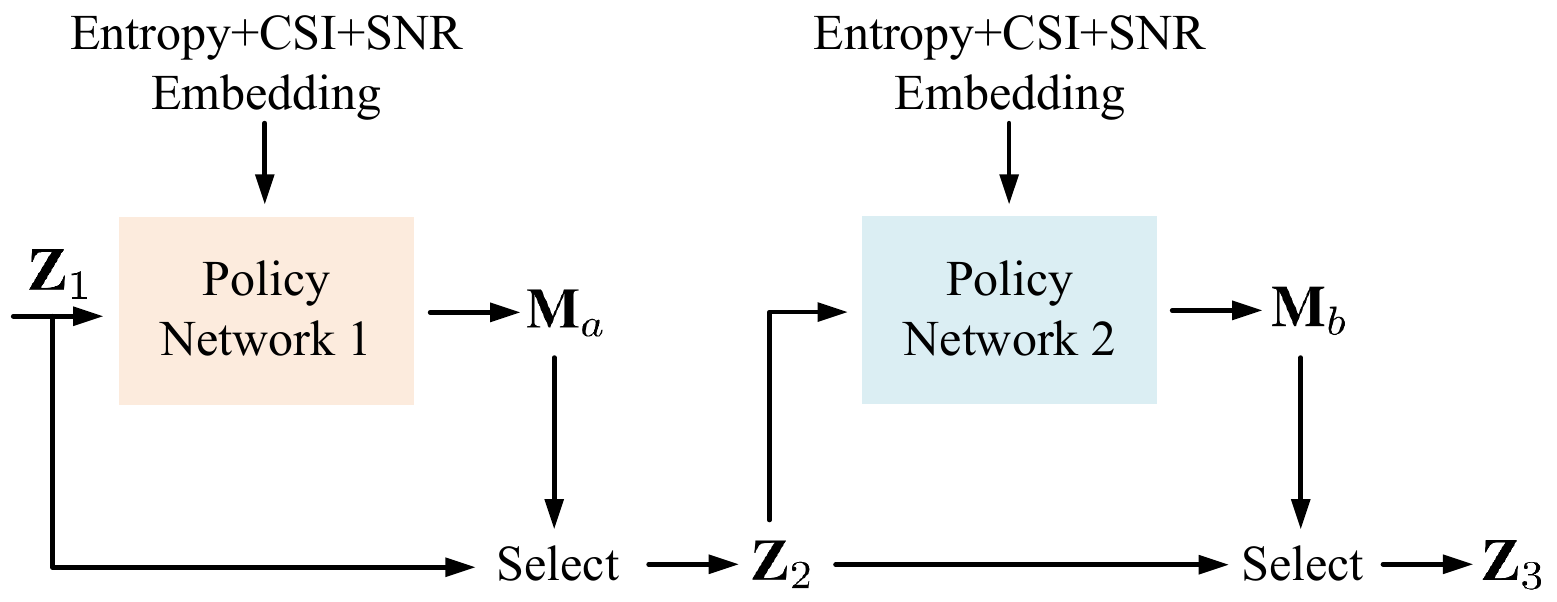}}
\caption{The network architecture of the joint feature map selection and pruning module.}
\label{fig.fmp}
\end{center}
\vskip -0.2in
\end{figure}

\begin{figure}[t]
\begin{center}
\centerline{\includegraphics[width=1\linewidth]{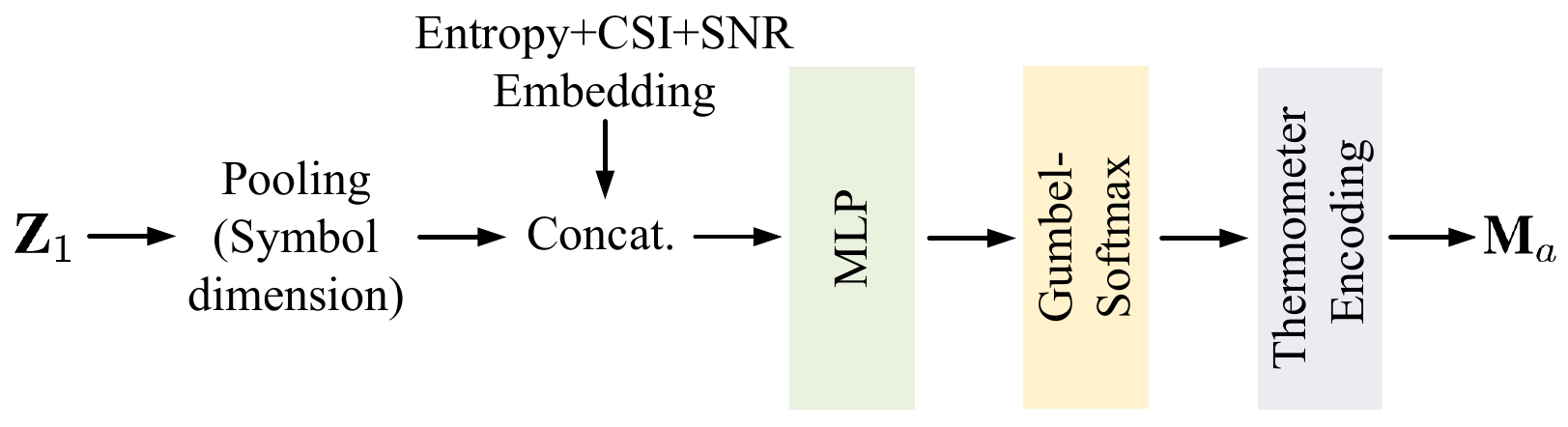}}
\caption{The network architecture of the policy network 1.}
\label{fig.policy1}
\end{center}
\vskip -0.2in
\end{figure}

\begin{figure}[t]
\begin{center}
\centerline{\includegraphics[width=1\linewidth]{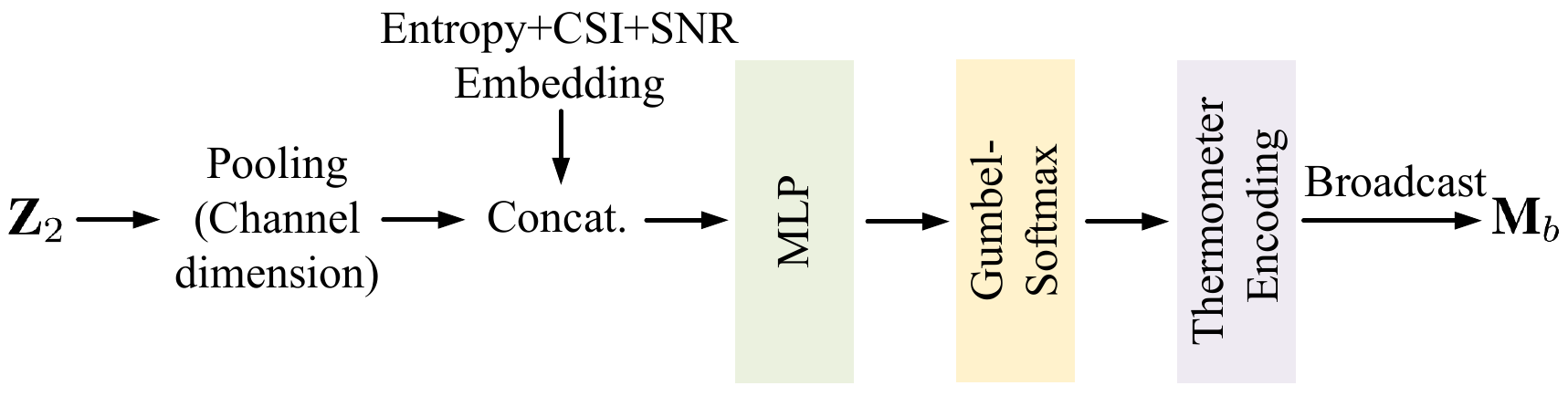}}
\caption{The network architecture of the policy network 2.}
\label{fig.policy2}
\end{center}
\vskip -0.2in
\end{figure}

As illustrated in Fig.~\ref{fig.fmp}, the proposed joint feature map selection and pruning module consists of a feature map selection module and a feature map pruning module, and operates on the feature maps $\textbf{Z}_{1}$ and the entropy-and-channel-condition embedding $\textbf{EC}_{\rm emb}$.
The goal of this module is to adaptively discard unimportant feature maps and prune redundant symbols within the retained feature maps, thereby achieving fine-grained, entropy-and-channel-aware adaptive rate control under varying channel conditions.
First, $\textbf{Z}_{1}$ and $\textbf{EC}_{\rm emb}$ are fed into the first policy network ($PN_1$). 
This network outputs a channel-wise binary mask $\textbf{M}_{a}$ that indicates which feature maps should be preserved.
By applying this mask to $\textbf{Z}_{1}$, we obtain a selected subset of feature maps, denoted by $\textbf{Z}_{2}$, where the discarded feature maps are set to zero.
Next, the selected feature maps $\textbf{Z}_{2}$ are processed together with $\textbf{EC}_{\rm emb}$ by the second policy network ($PN_2$).
This network produces a symbol-wise binary mask $\textbf{M}_{b}$ for the selected feature maps, specifying how many symbols within them should be preserved.
By applying this mask, we obtain the final pruned feature maps $\textbf{Z}_{3}$, where the pruned symbols are set to zero.

Overall, the joint module realizes a two-stage, entropy-and-channel-aware rate adaptation mechanism:
$PN_1$ performs coarse channel-wise selection, while $PN_2$ conducts fine-grained symbol-wise pruning.
The resulting pruned feature maps $\textbf{Z}_{3}$ are then normalized and transmitted over the MIMO Rayleigh fading channel.
It is worth noting that the masks $\textbf{M}_{a}$ and $\textbf{M}_{b}$ do not need to be transmitted to the receiver.
Since both masks follow a monotonic, prefix-preserving pattern, that is, each row is of the form $[1,\ldots,1,0,\ldots,0]$ along the channel or symbol dimension, the discarded feature maps and symbols always correspond to trailing zero positions. Hence, there is no ambiguity in their ordering, and the receiver can fully recover the structure of the feature maps from \textit{a single cut-off index} that specifies how many symbols per feature map are retained.
This is because the same symbol-wise cut-off is applied to all feature maps, and the retained symbols are concatenated in a predetermined order, by channel index.
Given the cut-off index, the receiver can partition the received symbol stream into equal-length groups for each feature map and zero-pad the missing (discarded) symbols to recover the original feature-map structure without any extra side information. 
According to \cite{2025swinjscc}, to ensure lossless transmission of the cut-off index, we adopt entropy coding. Since only one symbol needs to be additionally transmitted to the receiver per image, the corresponding bandwidth overhead is negligible and is therefore ignored in our rate calculation.
Next, we introduce the 2D entropy \cite{liu2023filter} and the two policy networks for feature map selection and pruning.


\subsubsection{2D Entropy} 

To quantify the information content of each feature map, we adopt the 2D entropy measure proposed in \cite{liu2023filter}.
Consider a latent representation with dimensions $H_{\rm f} \times W_{\rm f} \times C_{\rm f}$, where $C_{\rm f}$ is the number of feature maps (channels), while $H_{\rm f}$ and $W_{\rm f}$ denote the height and width of each feature map, respectively.
For the $i$-th feature map, denoted by $\textbf{FM}_i \in \mathbb{R}^{H_{\rm f} \times W_{\rm f}}$, its 2D entropy is computed as follows.
For each pixel in $\textbf{FM}_i$, let $m$ be its gray value and $n$ be the mean gray value of its local neighborhood.
By scanning over all spatial positions, we count how many times each pair $(m,n)$ appears in the feature map, denoted by $q(m,n)$.
This induces a joint probability distribution
\begin{equation}
    P_{m,n} = \frac{q(m,n)}{H_{\rm f} \times W_{\rm f}},
\end{equation}
where $P_{m,n}$ is the empirical probability that a pixel with gray value $m$ appears together with neighborhood mean $n$.
The 2D entropy of $\textbf{FM}_i$, denoted by $H(\textbf{FM}_i)$, is then defined as the Shannon entropy of this joint distribution:
\begin{equation}
    H(\textbf{FM}_i)
    = - \sum_{m,n} P_{m,n} \log_2 P_{m,n}.
\end{equation}
This 2D entropy simultaneously captures the intensity statistics and local spatial structure of the feature map, providing a comprehensive measure of its information richness.

\subsubsection{Policy Network 1--Feature Map Selection} 

Policy network 1 ($PN_1$) is designed to perform feature map selection, as illustrated in Fig.~\ref{fig.policy1}.
Given $\textbf{Z}_{1}\in\mathbb{R}^{L\times CU}$, $PN_1$ first applies a global average pooling over the symbol dimension to obtain a channel-wise summary vector in $\mathbb{R}^{CU}$, which is then concatenated with $\textbf{EC}_{\rm emb}$.
The concatenated vector is fed into a two-layer MLP to produce a $(CU+1)$-dimensional logit vector, where $CU$ is the number of feature maps.
Each entry of this logit vector corresponds to a candidate decision on how many feature maps should be preserved, i.e., from ``preserve none'' to ``preserve all''.
To obtain a discrete yet differentiable selection during training, we adopt the Gumbel-Softmax technique \cite{gumbelsoftmax} to sample an approximate one-hot decision from these logits, controlled by a temperature parameter.
Finally, the one-hot decision is converted into a channel-wise binary mask $\textbf{M}_{a}\in\{0,1\}^{CU}$ by thermometer encoding, so that $\textbf{M}_{a}$ takes the form $[1,\ldots,1,0,\ldots,0]$.
In other words, $PN_1$ decides a cut-off index for each sample, keeps the first few most important feature maps, and sets all subsequent feature maps to zero.

\subsubsection{Policy Network 2--Feature Map Pruning}

Policy network 2 ($PN_2$) is designed to perform feature map pruning, as illustrated in Fig.~\ref{fig.policy2}.
Given $\textbf{Z}_{2}\in\mathbb{R}^{L\times CU}$, $PN_2$ operates in a symbol-wise manner while enforcing a shared cut-off index across all feature maps within the same sample.
Concretely, a global average pooling is applied to $\textbf{Z}_{2}$ along the channel dimension to obtain an $L$-dimensional symbol-wise summary vector, which is then concatenated with $\textbf{EC}_{\rm emb}$.
The concatenated vector is fed into a two-layer MLP to produce a $(L+1)$-dimensional logit vector indicating how many symbols should be preserved.
Similar to $PN_1$, we apply the Gumbel-Softmax technique to obtain a one-hot decision from these logits, and then use thermometer encoding to convert this decision into a length-$L$ binary vector.
This yields a symbol-wise mask $\textbf{m}_{b}\in\{0,1\}^{L}$ of the form $[1,\ldots,1,0,\ldots,0]$, which specifies a single cut-off index along the symbol dimension.
The final mask is obtained by broadcasting $\textbf{m}_{b}$ to all feature maps, resulting in $\textbf{M}_{b}\in\{0,1\}^{L\times CU}$.

\subsection{InternViT-Based Feature Compensation Module}

\begin{figure}[t]
\begin{center}
\centerline{\includegraphics[width=0.9\linewidth]{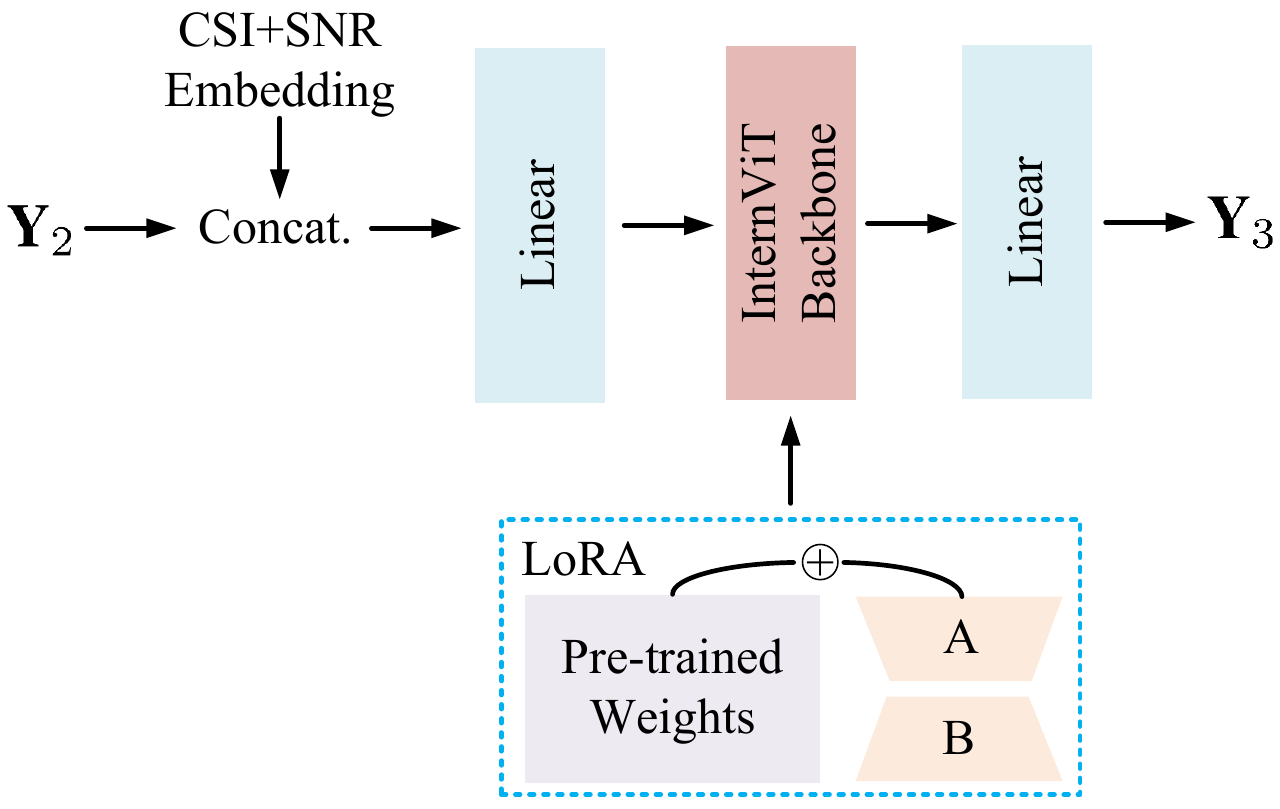}}
\caption{The network architecture of the InternViT-based feature compensation module.}
\label{fig.internvit}
\end{center}
\vskip -0.2in
\end{figure}


Modern MLLMs possess strong visual understanding capabilities \cite{2024mllm}.
Motivated by this, we exploit a powerful pre-trained MLLM backbone to compensate for the information loss caused by MIMO Rayleigh fading channels and joint feature map selection and pruning.
In particular, we employ the visual encoder (InternViT-300M) of the pre-trained InternVL3.5-1B model \cite{internvl35} developed by Shanghai AI Lab as the backbone of our feature compensation module, and the overall architecture of this module is illustrated in Fig.~\ref{fig.internvit}.
To adapt the pre-trained InternViT backbone to our latent feature space while keeping both training and inference costs manageable, we use a truncated InternViT consisting of the first half of its Transformer layers, insert it between two linear projection layers, and fine-tune it using LoRA \cite{lora2022}.
Concretely, the input to this module is $\textbf{Y}_2$, which is first concatenated with the channel-condition embedding $\hat{\textbf{CH}}_{\rm emb}$ along the channel dimension to incorporate channel information into the compensation process.
The concatenated features are then mapped by a linear projection layer to the hidden dimension of InternViT, so that they can be interpreted as token embeddings compatible with the vision Transformer (ViT) backbone.
The transformed tokens are subsequently fed into the truncated InternViT backbone, where the original backbone weights are frozen and only a set of low-rank LoRA adapters are trained within selected attention and MLP layers.
Finally, another linear projection layer maps the InternViT output back to the original channel dimension $CU$, yielding the refined feature maps $\textbf{Y}_{3}$.

\subsection{Channel-Aware Loss Function Design}

We train the proposed system in an end-to-end manner using a multi-objective loss that jointly accounts for reconstruction fidelity, channel usage, and feature consistency under varying channel conditions. 
During training, the channel SNR (in dB) of each sample, denoted by $\gamma$, is independently drawn from a uniform distribution
\begin{equation}
    \gamma \sim \mathcal{U}(\gamma_{\min}, \gamma_{\max}),
    \gamma_{\min}=0~\mathrm{dB},\ \gamma_{\max}=20~\mathrm{dB},
\end{equation}
so that the proposed model can learn to adapt to a wide range of channel conditions.

\subsubsection{Reconstruction Loss}

Let $\textbf{X}$ and $\hat{\textbf{X}}$ denote the source and reconstructed images, respectively.
We use the mean squared error (MSE) between $\textbf{X}$ and $\hat{\textbf{X}}$ as the reconstruction loss:
\begin{equation}
    \mathcal{L}_{\mathrm{rec}}
    = \mathrm{MSE}(\textbf{X}, \hat{\textbf{X}}).
\end{equation}

\subsubsection{Rate Regularization Loss}

As discussed previously, the joint feature map selection and pruning module produces a channel-wise mask $\textbf{M}_{a}\in\{0,1\}^{CU}$ and a symbol-wise mask $\textbf{m}_{b}\in\{0,1\}^{L}$.
Both masks follow a monotonic, prefix-preserving pattern of the form $[1,\ldots,1,0,\ldots,0]$, and $PN_2$ enforces that the same cut-off index along the symbol dimension is shared across all feature maps within a sample.

The effective binary mask $\textbf{M}\in\{0,1\}^{L\times CU}$ applied to the feature maps is defined element-wise as $\textbf{M}_{\ell,c}=\textbf{M}_{a,c}\,\textbf{m}_{b,\ell}$.
In practice, $\textbf{M}$ is obtained by broadcasting $\textbf{M}_{a}$ along the symbol dimension and $\textbf{m}_{b}$ along the channel dimension, such that $\textbf{M}_{\ell,c}=1$ if and only if $\textbf{M}_{a,c}=1$ and $\textbf{m}_{b,\ell}=1$.

%

Accordingly, the number of active (transmitted) symbols for one image is
\begin{equation}
    S_{a} = \sum_{\ell=1}^{L}\sum_{c=1}^{CU} \textbf{M}_{\ell,c}.
\end{equation}
The channel usage (compression ratio) of this sample is then
\begin{equation}
    CR = \frac{S_{a}}{2 \times 3 H W}.
\end{equation}

To make the rate penalty depend on the channel conditions, and thereby encourage the system to use more channel resources under poor channel conditions while saving resources when the channel conditions are good, we normalize $\gamma$ as
\begin{equation}
    \gamma_{\mathrm{norm}}
    = \frac{\gamma - \gamma_{\min}}{\gamma_{\max} - \gamma_{\min}} \in [0,1],
\end{equation}
and set a channel-aware weight
\begin{equation}
    \lambda_{\mathrm{rate}}(\gamma)
    = \lambda_{\mathrm{ch}}\cdot [\beta + (1-\beta)\,\gamma_{\mathrm{norm}} ],
\end{equation}
where $\lambda_{\mathrm{ch}}>0$ is a base channel-usage hyperparameter, and $\beta=0.6$.
The channel-aware rate regularization term is then
\begin{equation}
    \mathcal{L}_{\mathrm{rate}}
    = \lambda_{\mathrm{rate}}(\gamma)\cdot CR.
\end{equation}

\subsubsection{Feature Consistency Loss}

Let $\textbf{Z}_{1}\in\mathbb{R}^{L\times CU}$ denote the original feature maps produced by the semantic encoder, and let
$\textbf{Y}_{3}\in\mathbb{R}^{L\times CU}$ denote the refined feature maps obtained after MIMO Rayleigh fading channel transmission and InternViT-based feature compensation.
To encourage the feature compensation module to compensate for discarded features and denoise distorted ones, we introduce a feature consistency term:
\begin{equation}
    \mathcal{L}_{\mathrm{cons}}
    = \mathrm{MSE}(\textbf{Y}_{3}, \textbf{Z}_{1}).
\end{equation}

\subsubsection{Overall Objective}

The total channel-aware multi-objective training loss is given by
\begin{equation}
    \mathcal{L}_{\mathrm{total}}
    = \mathcal{L}_{\mathrm{rec}}
    + \lambda_{\mathrm{rate}}(\gamma)\cdot CR
    + \lambda_{\mathrm{cons}}\cdot\mathcal{L}_{\mathrm{cons}},
\end{equation}
where $\lambda_{\mathrm{cons}}>0$ controls the strength of the feature consistency regularization.
In summary, a larger $\lambda_{\mathrm{ch}}$ encourages more aggressive compression, resulting in a smaller CR, while a larger $\lambda_{\mathrm{cons}}$ encourages the refined features to be closer to the original ones.
Together, these terms guide the model to balance distortion, rate, and feature compensation under varying channel conditions.


\section{Performance Evaluation}


\subsection{Experimental Settings}

We use the CIFAR-10 dataset \cite{krizhevsky2009learning} for the source images, which consists of 50,000 training images and 10,000 testing images, all of which are $32\times 32\times 3$ RGB images.

We consider both $N_t \times N_r = 2 \times 2$ and $N_t \times N_r = 4 \times 4$ MIMO Rayleigh fading channels. The channel coefficients follow $h_{i,j} \sim \mathcal{CN}(0,\sigma_h^2)$ with $\sigma_h^2 = 1/N_t$. The AWGN has variance $\sigma_n^2$, which is set according to the target received SNR using Eq.~\eqref{snr-eq}. During training, for each sample, the SNR $\gamma$ (in dB) is independently drawn from a uniform distribution, i.e., $\gamma \sim \mathcal{U}(0,20)$. During testing, we fix $\gamma$ to a given value and report the average PSNR and CR over the entire test set. 
For both the proposed method and the benchmarks, we assume perfect estimates of the CSI and the SNR, and we do not account for the pilot transmission overhead in our experiments.

We adopt a two-stage SwinJSCC backbone for both the semantic encoder and decoder, applying stride-2 downsampling twice in the encoder and the corresponding upsampling in the decoder.
The patch size is set to $2\times 2$. The channel dimensions of the two encoder stages are set to 96 and 128, respectively. The first stage employs two channel-aware Swin Transformer blocks with six self-attention heads per block, while the second stage employs four blocks with eight heads per block. After the final stage, a linear projection layer maps the encoder output to a channel dimension $CU$, where we set $CU \in \{24, 36\}$ in our experiments.
The semantic decoder largely mirrors this structure in reverse. The two decoder stages use channel dimensions 128 and 96, respectively. The first decoder stage employs two channel-aware Swin Transformer blocks with eight self-attention heads per block, and the second stage employs four blocks with six heads per block.

Given this architecture, the feature maps produced by the semantic encoder have size $L \times CU$, where $L = \frac{H}{2^2}\cdot \frac{W}{2^2} = 64$.
Using the definition $CR = \frac{S}{2 \times 3HW}$, the maximum achievable CR for a given $CU$ is $CR_{\max} = \frac{L \times CU}{2 \times 3HW} = \frac{CU}{96}$. 
Therefore, with $CU = 24$ and $CU = 36$, the effective CR can adaptively vary within the ranges $[0, 0.25]$ and $[0, 0.375]$, respectively, depending on how many feature maps and symbols are selected by the two policy networks.
The length of the channel-condition embedding is set to 32, and the length of the EC embedding is set to 64.
Each policy network uses a two-layer MLP, and the temperature parameter of the Gumbel-Softmax is set to 5.

For the InternViT-based feature compensation module, to adapt the pre-trained vision encoder backbone to our task in a parameter-efficient manner, we apply LoRA \cite{lora2022} to both the self-attention and feed-forward sub-layers of each Transformer block, with rank $r=8$, scaling factor $\alpha=16$, and dropout rate 0.05.

For the channel-aware loss function, the rate regularization weight $\lambda_{\mathrm{ch}}$ is chosen from $\{100, 200\}$. A larger $\lambda_{\mathrm{ch}}$ imposes a stronger penalty on channel usage, encouraging more aggressive compression and thus a smaller average CR. The feature consistency weight $\lambda_{\mathrm{cons}}$ is fixed to $1\times 10^{-3}$. 
We train the model for 500 epochs with a batch size of 512 using the Adam optimizer and a learning rate of $1\times 10^{-4}$. 
All experiments are conducted on a single NVIDIA RTX A6000 GPU.

\subsection{The Benchmarks}

\subsubsection{BPG+LDPC}

We adopt the conventional separation-based source and channel coding scheme as the first benchmark, where BPG is used for image compression, LDPC codes are employed for channel coding, and quadrature amplitude modulation (QAM) is used for modulation. We refer to this benchmark as ``BPG+LDPC''.

In our simulations, the BPG encoder is implemented using the JCT-VC HEVC codec \cite{lainema2012intra}, and the color precision of each pixel is set to 8 bits. The channel coding stage follows the DVB-S2 LDPC standard with a coding rate of 1/2. 
For a fair comparison with the proposed system, we carefully match the effective CR. 
Specifically, for each (SNR, CR) pair achieved by our method, we select an appropriate quantization parameter for the BPG encoder together with a suitable QAM modulation order.
In this way, the BPG+LDPC scheme matches the CR of our system as closely as possible while keeping the LDPC code rate fixed.

\subsubsection{SwinJSCC+SA\&RA}

We adopt SwinJSCC with both SNR and rate adaptation modules \cite{2025swinjscc} as the second benchmark, denoted as ``SwinJSCC+SA\&RA''. This model represents the SOTA adaptive-rate SemCom system.
In our experiments, we extend SwinJSCC+SA\&RA to the same $N_t \times N_r$ MIMO Rayleigh fading setting as the proposed method, and employ the same L-MMSE detector at the receiver to ensure a fair comparison. The encoder and decoder architectures in this benchmark are configured identically to those in our proposed model.

Note that our proposed method only needs to transmit a single cut-off index to the receiver, whereas SwinJSCC+SA\&RA relies on transmitting an explicit binary mask over the channel dimension. Although the side-information overhead is ignored for both methods in our experiments, the cut-off index is inherently more compact, thus making our method practically more advantageous in terms of the achievable CR.

SwinJSCC+SA\&RA performs joint SNR and rate adaptation as follows. During training, for each sample, the SNR $\gamma$ (in dB) is drawn in the same way as in our method, and a target CR value is sampled from a predefined set of discrete CR values that uniformly span the same CR range as our method. At test time, both the SNR and target CR are fixed to specified values.
This benchmark is trained for 500 epochs with a batch size of 512 using the Adam optimizer with an initial learning rate of $1\times 10^{-4}$.

\subsection{Evaluation of Our Adaptive Rate Control Strategy}

\begin{table}[t]
\centering
\caption{Learned adaptive rate control strategy of the proposed method for $CU=36$, $N_t=N_r=2$, and $\lambda_{\mathrm{ch}}=100$: average compression ratio (CR) and PSNR versus SNR.}
\label{tab:strategy_lambda100}
\begin{tabular}{c|c|c|c|c|c}
\hline
SNR (dB)          & 0      & 5      & 10     & 15     & 20     \\
\hline
Average CR        & 0.2615 & 0.2230 & 0.2008 & 0.1943 & 0.1916 \\
\hline
Average PSNR (dB) & 23.81  & 24.91  & 25.41  & 25.94  & 26.11  \\
\hline
\end{tabular}
\end{table}

\begin{table}[t]
\centering
\caption{Learned adaptive rate control strategy of the proposed method for $CU=36$, $N_t=N_r=2$, and $\lambda_{\mathrm{ch}}=200$: average compression ratio (CR) and PSNR versus SNR.}
\label{tab:strategy_lambda200}
\begin{tabular}{c|c|c|c|c|c}
\hline
SNR (dB)          & 0      & 5      & 10     & 15     & 20     \\
\hline
Average CR        & 0.2008 & 0.1746 & 0.1614 & 0.1544 & 0.1490 \\
\hline
Average PSNR (dB) & 23.76  & 24.68  & 25.19  & 25.65  & 25.66  \\
\hline
\end{tabular}
\end{table}

In this subsection, we analyze the entropy-and-channel-aware adaptive rate control strategy learned by the proposed system. In particular, we investigate how the learned CR and the reconstruction quality (task performance), measured in terms of PSNR, vary with the channel SNR and with the rate regularization parameter $\lambda_{\mathrm{ch}}$, based on the numerical results reported in Tables~\ref{tab:strategy_lambda100} and~\ref{tab:strategy_lambda200}.

From Table~\ref{tab:strategy_lambda100}, which corresponds to $\lambda_{\mathrm{ch}} = 100$, we observe that the proposed system learns a reasonable rate-distortion behavior with respect to the channel SNR. When channel conditions are poor (SNR = 0 dB), the average CR is about 0.2615, and it gradually decreases to 0.1916 as the SNR increases to 20 dB. Meanwhile, the average PSNR increases monotonically from 23.81 dB to 26.11 dB. 
These results indicate that, under the proposed channel-aware loss function and with a fixed $\lambda_{\mathrm{ch}}$, the feature map selection and pruning modules learn to retain more feature maps and more symbols per feature map under harsh channel conditions, and to remove a substantial portion of symbols as the channel quality improves, while still enhancing task performance.

For the case with a stronger rate penalty, $\lambda_{\mathrm{ch}} = 200$, Table~\ref{tab:strategy_lambda200} reveals a more aggressive yet still smooth adaptive rate control behavior. Across all SNR points, the average CR is consistently lower than that in Table~\ref{tab:strategy_lambda100}. For example, at SNRs of 0 and 10 dB, the CRs are reduced from 0.2615 to 0.2008 and from 0.2008 to 0.1614, respectively, implying that only about 75\%-80\% of the symbols used under $\lambda_{\mathrm{ch}} = 100$ are transmitted.
Nevertheless, the corresponding PSNR values decrease only slightly. This indicates that, even with a much stronger penalty on channel usage, the proposed entropy-and-channel-aware mechanism can aggressively remove redundant features and symbols while preserving most of the information that is important for task performance.

Overall, these results confirm that, guided by the CSI, the SNR, the image content, the 2D entropy of the feature maps, and the proposed channel-aware loss function, the two policy networks can finely and effectively adjust the number of transmitted features and symbols. 
Accordingly, more communication resources are automatically devoted to poor channel conditions or to looser rate penalties (small $\lambda_{\mathrm{ch}}$), whereas redundant features and symbols are aggressively removed when channel conditions are good or when $\lambda_{\mathrm{ch}}$ is large.
Therefore, the proposed system demonstrates a strong capability to achieve an excellent rate-distortion tradeoff over a wide range of channel conditions.

\subsection{Performance Comparison with the Benchmarks}

\begin{figure}[t]
\centering
\subfigure[$\lambda_{\mathrm{ch}} = 100$]{
\begin{minipage}[t]{0.98\linewidth}
\centering
\includegraphics[width=1\linewidth]{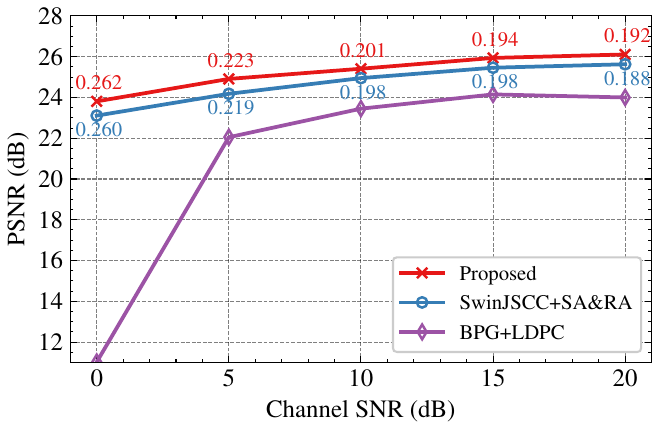}
\end{minipage}
}
\subfigure[$\lambda_{\mathrm{ch}} = 200$]{
\begin{minipage}[t]{0.98\linewidth}
\centering
\includegraphics[width=1\linewidth]{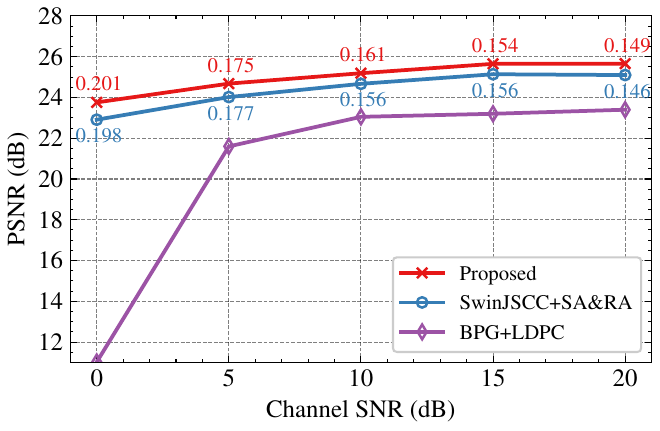}
\end{minipage}
}
\caption{Rate-distortion performance of the proposed system compared with the SwinJSCC+SA\&RA benchmark and the BPG+LDPC benchmark for $CU = 36$ and $N_t = N_r = 2$. Panels (a) and (b) correspond to $\lambda_{\mathrm{ch}} = 100$ and $\lambda_{\mathrm{ch}} = 200$, respectively. The CRs of each model at each operating point are labeled next to the corresponding curves, except for BPG+LDPC, since it shares the same CR as the proposed method.}
\label{fig:rd_CU36_Nt2}
\vskip -0.1in
\end{figure}

\begin{figure}[t]
\centering
\subfigure[$\lambda_{\mathrm{ch}} = 100$]{
\begin{minipage}[t]{0.98\linewidth}
\centering
\includegraphics[width=1\linewidth]{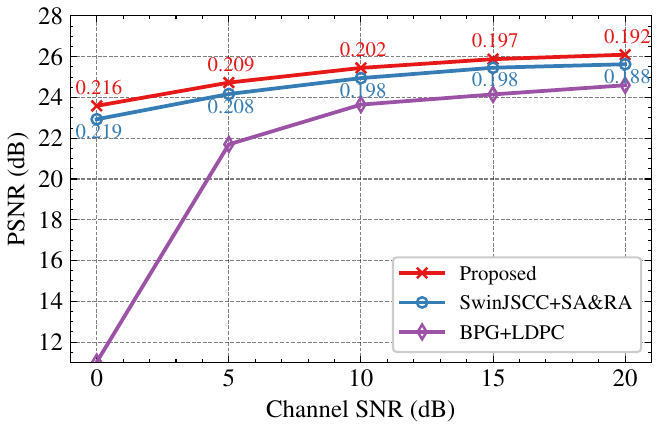}
\end{minipage}
}
\subfigure[$\lambda_{\mathrm{ch}} = 200$]{
\begin{minipage}[t]{0.98\linewidth}
\centering
\includegraphics[width=1\linewidth]{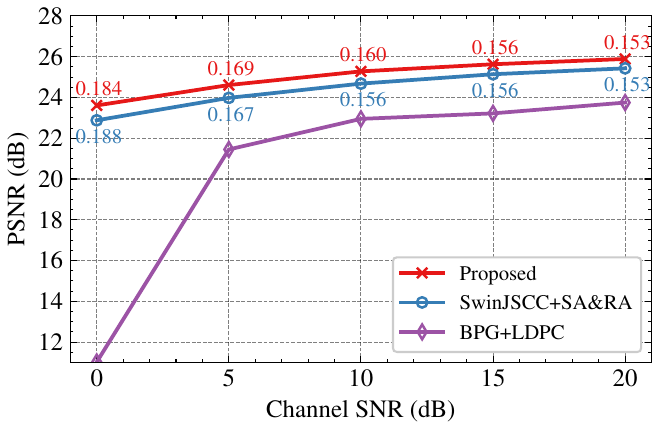}
\end{minipage}
}
\caption{Rate-distortion performance of the proposed system compared with the SwinJSCC+SA\&RA benchmark and the BPG+LDPC benchmark for $CU = 24$ and $N_t = N_r = 2$. Panels (a) and (b) correspond to $\lambda_{\mathrm{ch}} = 100$ and $\lambda_{\mathrm{ch}} = 200$, respectively. The CRs of each model at each operating point are labeled next to the corresponding curves, except for BPG+LDPC, since it shares the same CR as the proposed method.}
\label{fig:rd_CU24_Nt2}
\vskip -0.1in
\end{figure}

\begin{figure}[t]
\centering
\subfigure[$\lambda_{\mathrm{ch}} = 100$]{
\begin{minipage}[t]{0.98\linewidth}
\centering
\includegraphics[width=1\linewidth]{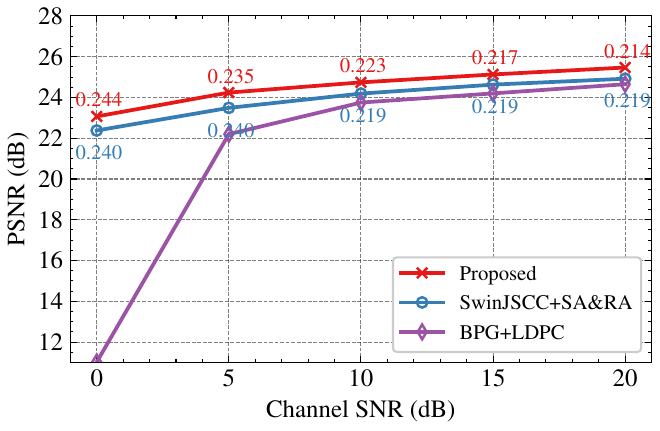}
\end{minipage}
}
\subfigure[$\lambda_{\mathrm{ch}} = 200$]{
\begin{minipage}[t]{0.98\linewidth}
\centering
\includegraphics[width=1\linewidth]{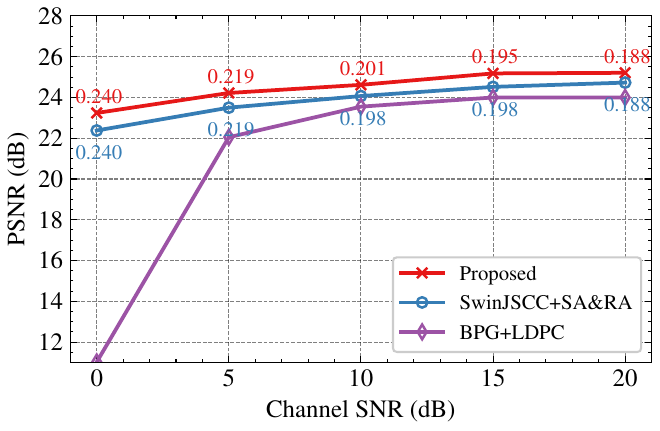}
\end{minipage}
}
\caption{Rate-distortion performance of the proposed system compared with the SwinJSCC+SA\&RA benchmark and the BPG+LDPC benchmark for $CU = 24$ and $N_t = N_r = 4$. Panels (a) and (b) correspond to $\lambda_{\mathrm{ch}} = 100$ and $\lambda_{\mathrm{ch}} = 200$, respectively. The CRs of each model at each operating point are labeled next to the corresponding curves, except for BPG+LDPC, since it shares the same CR as the proposed method.}
\label{fig:rd_CU24_Nt4}
\vskip -0.1in
\end{figure}

In this subsection, we compare the rate-distortion performance of the proposed system with the SwinJSCC+SA\&RA and BPG+LDPC benchmarks under different channel SNRs, rate regularization parameters $\lambda_{\mathrm{ch}}$, channel dimensions $CU$, and numbers of transmit and receive antennas ($N_t = N_r$), as illustrated in Figs.~\ref{fig:rd_CU36_Nt2}-\ref{fig:rd_CU24_Nt4}.
For the BPG+LDPC benchmark, the effective CR at each SNR operating point is exactly matched to that of the proposed system. 
For the SwinJSCC+SA\&RA benchmark, we set the target CR to be as close as possible to the average CR achieved by our proposed system at each SNR operating point.


We first consider $CU = 36$ and $N_t = N_r = 2$, as shown in Fig.~\ref{fig:rd_CU36_Nt2}. For $\lambda_{\mathrm{ch}} = 100$ in Fig.~\ref{fig:rd_CU36_Nt2}(a), the proposed system consistently achieves higher PSNR than SwinJSCC+SA\&RA while using nearly the same CR at each SNR. For instance, at SNRs of 5 dB and 10 dB, the proposed method reaches PSNRs of 24.91 dB and 25.41 dB, respectively, whereas SwinJSCC+SA\&RA attains only 24.18 dB and 24.95 dB. 
Compared with BPG+LDPC, in the medium-to-high SNR regime from 10 dB to 20 dB, the proposed method consistently outperforms BPG+LDPC by about 2 dB in PSNR at the same CR. When the rate penalty increases to $\lambda_{\mathrm{ch}} = 200$ in Fig.~\ref{fig:rd_CU36_Nt2}(b), all schemes move to lower CRs. The proposed method still maintains a clear rate-distortion advantage over both SwinJSCC+SA\&RA and BPG+LDPC.


Next, we consider $CU = 24$ and $N_t = N_r = 2$, as illustrated in Fig.~\ref{fig:rd_CU24_Nt2}. For $\lambda_{\mathrm{ch}} = 100$ in Fig.~\ref{fig:rd_CU24_Nt2}(a), the proposed system again lies above the SwinJSCC+SA\&RA curve in the rate-distortion plane. 
Across all SNRs, the CRs of the two schemes are extremely close, whereas the proposed method achieves a PSNR gain of about 0.4-0.7 dB. For example, at 0 dB the proposed method uses a slightly smaller CR of 0.216, while SwinJSCC+SA\&RA uses 0.219. In this case, the proposed method improves the PSNR from 22.93 dB to 23.59 dB. 
Compared with BPG+LDPC at the same CR, the proposed system provides more than 1.5 dB PSNR gain at SNRs above 10 dB.
When $\lambda_{\mathrm{ch}} = 200$ in Fig.~\ref{fig:rd_CU24_Nt2}(b), all methods operate at lower CRs. The proposed system still offers about 0.5-0.7 dB PSNR improvement over SwinJSCC+SA\&RA with only marginal CR differences, and it outperforms BPG+LDPC by roughly 2 dB in PSNR at medium-to-high SNRs.


Finally, Fig.~\ref{fig:rd_CU24_Nt4} reports the results for $CU = 24$ and a larger MIMO configuration with $N_t = N_r = 4$. For $\lambda_{\mathrm{ch}} = 100$ in Fig.~\ref{fig:rd_CU24_Nt4}(a), the proposed system consistently achieves higher PSNR than SwinJSCC+SA\&RA across all SNRs.
At low SNRs, this gain is achieved with only a small increase in CR. At medium-to-high SNRs, the proposed method can even achieve both higher PSNR and lower CR. For instance, at 20 dB SNR it attains 25.47 dB PSNR with a CR of 0.214, whereas SwinJSCC+SA\&RA reaches only 24.92 dB PSNR with a larger CR of 0.219, corresponding to a better rate-distortion operating point. Compared with BPG+LDPC at the same CR, the proposed system exhibits a clear advantage in PSNR.
When $\lambda_{\mathrm{ch}} = 200$ in Fig.~\ref{fig:rd_CU24_Nt4}(b), the superiority of the proposed system persists.
It improves PSNR by roughly 0.5-0.9 dB relative to SwinJSCC+SA\&RA at similar CRs. At some SNR values, such as 15 dB, it simultaneously uses a lower CR and yields noticeably higher PSNR. At the same CR, the proposed method also consistently outperforms BPG+LDPC.

Overall, across all considered configurations and both values of $\lambda_{\mathrm{ch}}$, the proposed system achieves higher PSNR at nearly the same or even lower CR compared with SwinJSCC+SA\&RA and BPG+LDPC.
These results demonstrate a consistently superior rate-distortion tradeoff over a wide range of channel conditions.

\subsection{Ablation Study}

\begin{figure}[t]
\centering
\subfigure[$\lambda_{\mathrm{ch}} = 100$]{
\begin{minipage}[t]{0.98\linewidth}
\centering
\includegraphics[width=1\linewidth]{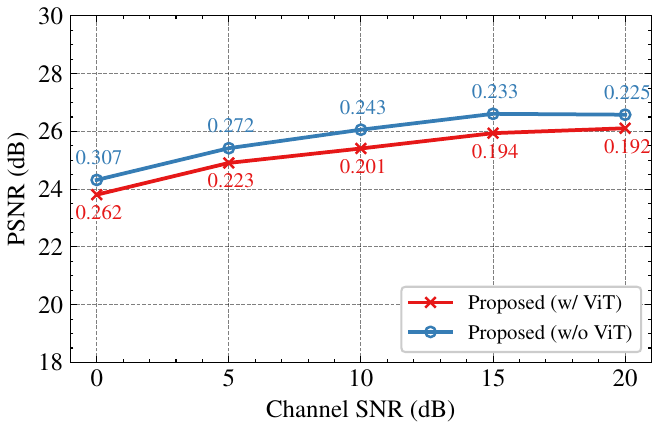}
\end{minipage}
}
\subfigure[$\lambda_{\mathrm{ch}} = 200$]{
\begin{minipage}[t]{0.98\linewidth}
\centering
\includegraphics[width=1\linewidth]{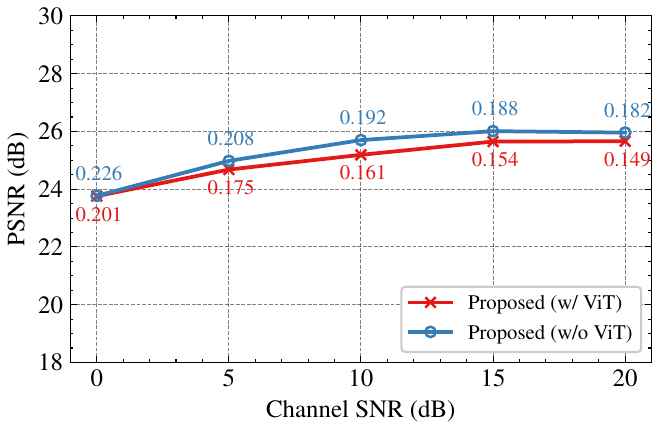}
\end{minipage}
}
\caption{Ablation study of the proposed method with (w/) and without (w/o) the InternViT-based feature compensation module for $CU = 36$ and $N_t = N_r = 2$, denoted as ``w/ ViT'' and ``w/o ViT'', respectively. Panels (a) and (b) correspond to $\lambda_{\mathrm{ch}} = 100$ and $\lambda_{\mathrm{ch}} = 200$, respectively. The CRs of each model at each operating point are labeled next to the corresponding curves.}
\label{fig:ablation_CU36_Nt2}
\vskip -0.1in
\end{figure}

In this subsection, we investigate the contribution of the InternViT-based feature compensation module through an ablation study. We compare the full model that includes this module, denoted as ``w/ ViT'', with a variant that excludes it, denoted as ``w/o ViT'', while keeping all other components and training settings unchanged. 
The rate-distortion performance for $CU = 36$ and $N_t = N_r = 2$ is shown in Fig.~\ref{fig:ablation_CU36_Nt2} for $\lambda_{\mathrm{ch}} = 100$ and $\lambda_{\mathrm{ch}} = 200$.

From Fig.~\ref{fig:ablation_CU36_Nt2}, we observe that for both $\lambda_{\mathrm{ch}} = 100$ and $\lambda_{\mathrm{ch}} = 200$, the model w/ ViT operates at significantly lower CRs than the model w/o ViT while maintaining very similar PSNR. For $\lambda_{\mathrm{ch}} = 100$, the CR decreases from 0.307-0.225 (w/o ViT) to 0.262-0.192 (w/ ViT), corresponding to roughly 15\%-18\% fewer transmitted symbols, while the PSNR is only about 0.5-0.7 dB lower across all SNRs. For $\lambda_{\mathrm{ch}} = 200$, the CR further drops from 0.226-0.182 (w/o ViT) to 0.201-0.149 (w/ ViT), i.e., by about 11\%-18\%, and the PSNR loss becomes even smaller.
In this case, the model w/ ViT achieves nearly the same PSNR at a substantially lower CR, thereby saving channel resources without noticeably compromising task performance.
For example, at 0 dB SNR, the model w/ ViT attains essentially the same PSNR as the model w/o ViT while using a clearly lower CR.

Overall, the ablation study demonstrates that the InternViT-based feature compensation module plays a key role in improving the rate-distortion performance of the proposed system.
By exploiting global contextual dependencies among the retained features and symbols, this module can effectively compensate for part of the information loss caused by channel distortion as well as feature map selection and pruning.
Consequently, the policy networks can learn more aggressive yet reliable adaptive rate control strategies, leading to substantially lower CRs with only marginal or even no PSNR degradation.

\section{Conclusion}

In this paper, we proposed a novel SemCom framework with entropy-and-channel-aware adaptive rate control over MIMO Rayleigh fading channels.
To realize this framework, we embedded a joint representation of the CSI and the SNR into both the semantic encoder and decoder by equipping them with channel condition adaptive modules, so that the feature maps are modulated according to varying channel conditions.
On top of this channel-aware architecture, we designed two policy networks to realize fine-grained joint feature map selection and pruning. The first policy network adaptively retains task-relevant feature maps, and the second prunes semantically redundant symbols within the selected feature maps, exploiting the fact that even highly informative feature maps still contain symbol-level redundancy. Driven jointly by the feature maps, their 2D entropy, the CSI, and the SNR, these policy networks achieve entropy-and-channel-aware rate adaptation. 
To compensate for information loss due to MIMO Rayleigh fading channels as well as feature map selection and pruning, we further employed a lightweight vision encoder InternViT-300M as an MLLM-aided feature compensation module. 
We used a truncated version of InternViT and fine-tuned it efficiently via LoRA, thereby reducing training and inference overhead. In addition, we designed a channel-aware loss function that encourages the system to allocate more resources under poor channels while saving resources under favorable channels, and at the same time maintains high task performance.
Extensive experiments demonstrated that the proposed system consistently outperforms conventional separation-based source and channel coding and SOTA adaptive-rate SemCom benchmarks in terms of rate-distortion performance.
%
Looking forward, we aim to extend the proposed framework to multi-user scenarios, where the transmitter performs differentiated adaptive transmission for multiple receiver groups with heterogeneous task objectives, channel conditions, rate constraints, and potentially distinct private knowledge bases.

\bibliographystyle{IEEEtran}

\bibliography{IEEEabrv,myref}
 
\vspace{12pt}

\end{document}